# Just AI Governance and the Forbidden-Boundary Principle

An Integrated Four-Layer Framework and an Auditable First-Principle Safe-Stop Protocol (FPSSP)

Na Young Ahn | *AI-assisted drafting disclosed below*

## Abstract

Artificial intelligence (AI) increasingly shapes consequential decisions, yet predictive accuracy is not a sufficient justification for delegated authority. This conceptual study develops an integrated framework for assessing just AI across normative, data, model, and governance layers. Drawing on plural theories of justice, sociotechnical fairness research, and publicly available governance standards, it maps six justice requirements to observable evidence, decision rights, and remedies. It proposes a two-stage evaluation rule: exclude uses that fail independently justified minimum rights constraints, then compare eligible alternatives across multiple performance and governance criteria. As an explicitly limited normative analogy, Genesis 2:15–17 motivates a Forbidden-Boundary Principle (FBP), under which an AI agent must not execute externally consequential actions outside pre-authorized boundaries. The resulting First-Principle Safe-Stop Protocol (FPSSP) comprises an independently governed policy monitor, a mandatory execution gate, a restricted or alternative safe-state controller, immutable-oriented audit trails, and accountable human authorization for recovery. We distinguish this normative and traceability contribution from existing Simplex and runtime assurance architectures; we do not claim a novel safety switch or formal proof of universal ethical compliance. A policy schema, threat model, test matrix, outcome definitions, and a prospective comparison design make the proposal falsifiable. No FPSSP implementation, experimental dataset, numerical improvement, or verified latency bound is reported. The framework supports future evaluation rather than constituting evidence of operational safety.

Keywords: just AI; algorithmic fairness; normative constraints; Forbidden-Boundary Principle; runtime assurance; execution gate; human oversight; safe state; auditability.

## 1. Introduction

Artificial intelligence is no longer limited to the role of a computational tool that simply executes human commands. AI is now involved in numerous decision-making processes that impact people's lives, ranging from loan screening, insurance pricing, recruitment, medical diagnostic assistance, personalized education, crime risk prediction, welfare eligibility, content recommendations, and patent and legal document analysis, to even military or security support. These decisions may take the form of mathematical models, statistical correlations, optimization functions, or probabilities. However, their consequences can affect reputation, livelihoods, safety, rights, and freedoms. For this reason, AI is viewed not merely as a high-performance tool, but as part of society's decision-making infrastructure.

It is precisely at this point that the issue of justice becomes important. In human society, justice means that people are treated as human beings, that each person receives what is due to them, and that power and resources are distributed according to just causes. Justice is not merely a legal rule or a moral slogan; it is a standard for evaluating social institutions. If AI systems now regulate opportunities and rights in recruitment, education, lending, welfare, healthcare, trials, and access to information, then such systems also need to be reviewed by the standards of justice. This principle must be applied not only at the final decision-making stage but also throughout the design, learning, deployment, operation, and post-deployment review processes.

The expression "just AI" can lead to several misunderstandings. One misconception is that justice can be reduced to a single mathematical objective function. Justice encompasses multiple domains and cannot be compressed into a single entity. Another misconception is that AI is more neutral than human decision-makers. In reality, AI can reproduce historical biases embedded in data, the assumptions of its designers, organizational interests, and the power relations of the usage environment. A third misconception is that merely writing ethical principles in a manual is sufficient. Justice typically operates not only through declarations but also through procedures, institutions, accountability, and remedies.

This paper makes three bounded contributions. First, it connects plural conceptions of justice to a four-layer requirements-and-evidence framework while separating non-compensatory rights conditions from compensatory performance trade-offs. Second, it specifies the Forbidden-Boundary Principle as a policy-development rule and situates a First-Principle Safe-Stop Protocol (FPSSP) within, rather than outside, established runtime assurance work. Third, it offers auditable policy fields, a concrete threat model, and prospective test procedures that distinguish assertions of design intent from measured engineering properties. These are conceptual and methodological contributions, not claims of implemented safety performance.

The methodology of this study combines literature-based normative analysis with conceptual design. This research examines major theories of justice and relates them to AI ethics, algorithmic fairness, international principles, and risk management frameworks. For the purposes of this study, justice is reconstructed into six elements: human dignity, procedural fairness, distributive fairness, recognition and non-discrimination, accountability and remedies, and communal solidarity. These elements are then mapped to the lifecycle of AI systems to propose design principles, evaluation metrics, and operational procedures.

Sections 2–5 establish the philosophical and sociotechnical motivation. Section 6 presents the four-layer model, the Genesis-based normative analogy, and the boundary-control architecture; new Sections 6.8–6.10 distinguish prior art, define executable obligations, and specify a threat model. Sections 7–9 address fairness, accountability, and field-specific scenarios. Section 10 sets out prospective assessment and falsification procedures; Section 11 identifies limitations; and Section 12 concludes. Reference metadata checks are described separately from experimental validation.

## 1.1. Research Questions and Academic Contributions

The central research questions of this study are as follows: First, what intermediate concepts and decision-making procedures are required to translate competing theories of justice into design requirements for AI systems? Second, how should technical verification regarding statistical fairness be combined with human dignity, the right to participation, and the right to object? Third, how can the limitations of an approach that evaluates AIs with varying risk levels and application fields using a single fixed score be overcome? Fourth, what is the minimum operational unit that transforms declarative principles into auditable evidence? These questions focus on the social distribution of performance and the justification structure of decision-making rather than the performance of the model itself. Fifth, how can the attempt and execution of serious acts exceeding the authority of an AI be separated and detected, and how can they be safely terminated under human control without amplifying damage that has already occurred?

This study corresponds the six elements of justice to four levels—norms, data, models, and governance—while distinguishing between the minimum obligations that must be met independently at each level and the obligations of linkage between levels. Through this, it connects outcome-oriented fairness measurement, procedure-oriented control, and ex post facto accountability and remedies into a single evaluation structure. This model does not claim to be the first to mathematize justice. Rather, it aims for a conceptual contribution that organizes existing philosophical discussions and research on fairness and accountability into questions and evidence that can be examined in practice (Selbst et al., 2019; Raji et al., 2020).

## 1.2. Method, Evidence Scope, and Limits of Verification

The study employs normative conceptual analysis, comparison of justice theories, synthesis of sociotechnical accountability research, and design-oriented specification. Primary publication and institutional records were consulted for core runtime assurance, algorithmic fairness, AI risk governance, and legal sources. These checks establish publication identity and support for specific propositions, not a systematic literature review, a legal opinion, or empirical confirmation of FPSSP. The bibliography is documented with persistent identifiers where verified. Research propositions are classified as (E1) established source findings, (E2) this paper's normative argument, (E3) hypothetical engineering requirements, and (E4) empirical claims requiring future tests. Theoretical compatibility does not imply an observed safety improvement.

No corpus of deployed AI systems, field participants, executable FPSSP prototype, safety-case verification result, or statistical comparison was generated for this manuscript. Examples, equations, tables, and numerical illustrations are explanatory. Reported properties of existing approaches are attributed to their corresponding publications and shall not be transferred to FPSSP without a new proof or measured experiment. This explicit boundary is essential to evaluating the contribution of a conceptual manuscript fairly.

## 1.3. Reference Verification and Claim-to-Source Discipline

A selected bibliographic cross-check was performed against the original publisher, institutional, or regulatory pages available at revision. Claims were separated into those concerning the existence of a publication, those actually supported by its stated research findings, and new hypotheses introduced by this manuscript. The check confirms, for example, the identity and limited scope of Abate et al. (2019), Slagel et al. (2024), Hardt et al. (2016), Chouldechova (2017), and the voluntary institutional role of NIST AI RMF (Tabassi, 2023). It does not validate every philosophical interpretation or establish operational effectiveness of FPSSP.

For reproducibility, an external reference-verification report associated with this revision records the verification date, authoritative location, correction, and unresolved items. Unverified bibliographic details must be checked against the final venue or publisher prior to journal submission. Source checking is not a substitute for a systematic literature review; the paper does not represent it as one.

## 2. The Meaning of Justice in Human Society

Justice is one of the oldest questions that human society has raised to justify its order. People ask why they must follow certain laws, why some people possess more resources than others, why the state can restrict freedom, why wrongdoing must be punished, and why the vulnerable deserve special protection. The concept of justice lies at the center of these questions. It asserts that social order must not depend solely on power or convenience, but must be based on reasons that can be justified.

The classic answer to the question of justice is "giving each person what is due." This phrase is simple, yet it immediately raises a difficult question: should what is due be determined by ability, need, legal rights, contribution, or the equal value of all human beings? Educational opportunities provide a familiar example. Some societies may base their criteria solely on academic performance, while others may consider economic disadvantage and regional inequality together. When AI is used for such allocation, the selection of criteria is not merely a technical setup; it is a judgment regarding justice.

Justice also has a procedural dimension. Even if an outcome appears useful or even correct, it is still not just if the process is unfair. Similar matters must be handled according to similar standards. Decision-making criteria should be disclosed as much as possible, and those affected must be able to express their opinions and challenge errors. This is why procedures are important in trials, administrative dispositions, patent examinations, recruitment, testing, and welfare selection. If procedures are opaque, it is difficult for affected parties to accept an outcome as legitimate, even if it is substantively correct. Therefore, the opacity of AI is not merely a technical issue, but also a matter of justice.

Justice is also a matter of distribution. Every society must decide how to allocate scarce resources and burdens. The subjects of distribution include not only income and goods, but also education, healthcare, information, political influence, cultural recognition, safety, time, and environmental burdens. Distributive justice is not the same as simple arithmetic equality. In some situations, equal treatment is required, while in others, greater support is needed for those in disadvantageous positions. In AI-based allocation, outcomes can vary significantly depending on whether the system optimizes cost, efficiency, inclusion, or protection of the vulnerable.

Humans are not merely objects of resources, but subjects whose identity and dignity must be respected. Discrimination cannot be explained solely by material deprivation. Stereotypes, the invisibility of language and culture, and exclusion based on disability, age, gender, or region can also deny recognition. Since AI models learn from image, language, voice, and behavioral data, they can determine which human experiences are treated as normal and which as marginal. That determination is directly related to justice.

Accountability and remedies are also essential. If harm results from a wrongful decision, someone must explain what happened, accept responsibility, and provide a pathway for correction. Justice does not stop at prohibiting wrongdoing; it is also a structure that ensures victims are not left without remedies. If an algorithm unfairly rejects a loan, disqualifies an applicant, assigns inaccurate risk scores, or generates discriminatory recommendations, the affected individual must be able to know the reason, file a complaint, receive a correction, and, if necessary, be compensated. Automation without accountability can become a highly efficient form of domination.

Finally, justice is related to the restoration of communal trust. Justice is not revenge. Justice seeks to identify harm, rectify wrongs, prevent recurrence, and restore trust within the community. When AI amplifies hate speech, misinformation, social conflict, or political polarization, it is not enough to simply ask whether individual predictions were accurate. Just AI systems must be designed so as not to unnecessarily undermine social trust and must consider vulnerable people and future generations.

### 2.1. Legitimacy of Outcome, Legitimacy of Procedure, Legitimacy of Structure

When applying the concept of justice to AI, it is necessary to distinguish between distinct units of analysis: outcome, procedure, and structure. The legitimacy of outcome concerns how a particular decision allocated opportunities and costs among people. The legitimacy of procedure concerns whether the criteria leading to that decision were disclosed, whether the perspectives of relevant parties were reflected, and whether objections were possible. The legitimacy of structure asks who sets the system's objectives and which groups have a voice in the model development process. These three units overlap but are not reducible to one another.

For example, even if an AI evaluating job applicants matches the hiring recommendation ratios for each group, procedural and structural issues remain if there is no way for applicants to correct the results and if individuals with disabilities are blocked from participating in the model development. Conversely, even if the process is transparent and user-friendly, the legitimacy of the results cannot be secured if a specific group experiences unfavorable outcomes over a long period. Therefore, the question of justice evaluation must include not only "are the results the same?" but also "who set the standards" and "is there a system to correct errors?"

### 2.2. Distinction between Formal Equality and Substantive Equality

Formal equality emphasizes applying the same rules to everyone. However, if the starting conditions under which the rules are applied differ significantly, the same rules can perpetuate inequality. Substantive equality does not imply the uniform homogenization of outcomes; rather, it focuses on removing obstacles that hinder institutional participation and ensuring each individual's actual potential for choice (Sen, 2009; Nussbaum, 2011). This is why, when the accurate classification of AI conflicts with the expansion of human capabilities, the system cannot be justified solely by the accuracy of classification.

Design measures to achieve substantive equality include accessible interfaces, multilingual descriptions, reasonable accommodations, sufficient human counseling, and reduced appeal costs. However, since these measures require prioritization within limited resources, public discussion is necessary regarding who should receive what additional support. Technical adjustments can support this public judgment but cannot substitute for its legitimacy.

### 2.3. Temporal and Intergenerational Dimensions of Justice

AI can alter the structure of future opportunities by repeatedly applying the same decisions. If a predictive model allocates fewer resources to students given low learning opportunities, a cycle may occur where low achievement reinforces low predictions. Even if the current error rate is low, it is difficult to detect long-term injustices without evaluating this dynamic feedback. Therefore, the unit of evaluation for justice must encompass both the single decision and the entire distribution period.

When computational resource and energy consumption, data labor, and environmental burdens are added to this, the scope of justice extends to future generations and data producers. Since not all AI has the same environmental impact, harm cannot be determined without specific estimates. However, explicitly stating that the costs and benefits of large-scale systems may accrue to different groups, and documenting the structure of burden attribution, serves as a practical starting point for communal solidarity.

## 3. Philosophical Traditions of Justice and Their Implications for AI

Plato understood justice as the harmonious order of the soul and the city. From this perspective, justice exists when each part performs its unique role and the whole maintains balance. This idea can have implications for the design of modern AI systems, as many actors are involved, including data collectors, developers, distributors, users, supervisory bodies, and those affected. If any actor evades responsibility or shifts the role of another, the justice of the entire system can collapse. At the same time, since the theory of harmony can easily justify hierarchy, it must be combined with democratic participation and power control in today's context.

Aristotle distinguished between general justice and particular justice, and further divided particular justice into distributive justice and corrective justice. Distributive justice concerns the distribution of honor, wealth, and opportunities. Corrective justice concerns the recovery of unjust gains and losses. In AI, distributive justice manifests when a model allocates opportunities or resources. Corrective justice manifests when damage caused by automated decisions must be remedied. This distinction is useful because it demonstrates that just AI requires both pre-design and post-correction.

Kantian ethics centers on the idea that human beings should never be treated merely as means, but always simultaneously as ends. This principle holds special significance in the data economy. Human behavior, preferences, emotions, relationships, location, and health information are often transformed into learning resources for model improvement. In this process, humans can be treated as raw materials for prediction and optimization. From a Kantian perspective, just AI must not view humans solely as objects of prediction, manipulation, or classification. It must treat humans as autonomous subjects capable of demanding explanations, consenting to or rejecting, and requesting corrections. Therefore, human dignity is a limit that takes precedence over efficiency.

Utilitarianism pursues the greatest happiness for the greatest number. Because AI can process large-scale data and optimize large-scale outcomes, it can support utilitarian reasoning. This can be valuable in the allocation of medical resources, traffic control, energy conservation, and the selection of welfare recipients. However, utilitarian optimization can justify the sacrifice of minority rights. If higher error rates are allowed for small groups to increase overall accuracy, or if care is reduced for high-cost individuals, the outcome may be efficient but may not be just. Therefore, AI aiming for justice must set limits on utility maximization when fundamental rights are at issue.

Rawls's theory of justice remains significant in contemporary discourse. Rawls viewed justice as the primary virtue of social institutions and sought to derive fair principles through the original position and the veil of ignorance. His two principles are generally summarized as equal basic liberties and the difference principle. According to the difference principle, social and economic inequalities are justified only when they benefit the most disadvantaged. Applied to AI, the Rawlsian approach requires evaluating not only average performance but also the impact on the most disadvantaged groups. Efficiency cannot justify AI practices that undermine basic rights and liberties.

Sen and Nussbaum's capabilities approach argues that the quality of human life cannot be judged solely by the quantity of distributed resources. What matters is what people can actually do and what kind of life they can live. Even if the same educational services are provided, actual capabilities can vary depending on disability, language, poverty, caregiving burdens, or regional disparities. When AI is used in education or welfare, it must not stop at providing the same interface for everyone. It must expand actual accessibility and available capabilities. This shifts the discussion from formal equality to substantive opportunity.

Fraser's explanation is useful in that it addresses justice through distribution, recognition, and representation. Distribution concerns economic resources. Recognition concerns cultural respect and discrimination. Representation concerns who can participate in the decision-making process. All three dimensions appear in AI governance. Who is included in datasets, whose performance is measured, and whether affected people can participate in design and evaluation are all questions of justice. Therefore, just AI must not stop at reducing biased outcomes but must also improve the representativeness of the system-building process.

Care ethics and communitarian thought remind us that justice cannot be explained solely by abstract individual rights. Humans live within dependence, care, trust, and relationships. If AI renders care labor invisible in the name of efficiency or treats human relationships merely as a cost, it misses a crucial moral dimension. In elder care, disability support, child education, and mental health counseling, AI should support human relationships rather than replace them. Otherwise, even if services appear efficient, vulnerable individuals may become further isolated.

**Table 1. Key Justice Theories and Implications for AI Design**

| Theory of Justice | Key Concepts | Relevance to AI Design |
|---|---|---|
| Plato and Aristotle | Harmony, roles, distribution, and correction | Responsibility by Actor; Damage Recovery Procedures |
| Kant | Treat humans as ends in themselves | Protection of dignity, autonomy, consent, and the right to rectification |
| Utilitarianism | Total utility maximization | Optimization considering restrictions on minority rights |
| Rawls | Basic liberties, difference principle | Impact Assessment on Vulnerable Groups; Guarantee of Fundamental Rights |
| Sen and Nussbaum | Substantive capabilities | Evaluation of actual accessibility and selectability |
| Fraser | Distribution, recognition, representation | Data Representativeness; Participatory Design |
| Care ethics | Relationships and Vulnerability | AI Design to Support Human Relationships |

## 3.1. Conflicts and Priority Ordering among Theories of Justice

Design principles are not determined merely by arranging philosophical theories of justice into a single list. Rawls's demand for basic liberties places restrictions on choices that sacrifice the rights of minorities for the benefit of the majority. Utilitarianism excels at comparing the total benefits resulting from multiple options. The capabilities approach focuses on whether people provided with the same resources can actually use them. Fraser's perspective reveals who holds the representation to participate in decision-making and the structural combination of distribution and recognition (Rawls, 1971; Sen, 2009; Fraser, 2008).

These perspectives are complementary but not always harmonious. For instance, regarding the question of whether it is acceptable to increase the error rate of specific minority groups for the sake of high accuracy, the logic of efficiency can conflict with rights-based restrictions. Just AI governance should not conceal such conflicts as a single weighted sum; instead, it must first identify the options to be excluded and then perform a comparison among acceptable alternatives. The 'non-compensatory lower bound on rights' proposed in this study is a design principle that does not offset significant threats to dignity and fundamental rights with average benefits. This is not a mere reproduction of the direct norms of specific laws, but rather a proposal derived from various theories of justice.

## 3.2. Translation from Philosophical Principles to Design Questions

The practical unit for applying justice theory is not abstract values, but answerable questions. Dignity is transformed into the question, "Is an individual treated merely as a score through this system?", procedural justice into, "Do decision-makers have access to the rationale and rectification pathways?", and corrective justice into, "If a wrongful decision is identified, to whom and what restitution can be demanded?" Distributive justice can be concretized as, "To which groups are errors and benefits concentrated?", recognition justice as, "Have there been experiences of omission or misrepresentation?", and representation justice as, "Did affected parties participate in design decision-making?"

These questions do not imply a complete measurement of philosophical concepts. Each serves as a mechanism to request observable evidence, such as documents, meeting minutes, audit results, performance sheets, interviews, and relief records. Particularly in cases where quantitative data is absent, stakeholder experience and independent qualitative review must be utilized in conjunction. The purpose of this translation work is to ensure that the verifiability provided by technical evaluation and the context provided by humanities and social science interpretations limit and complement each other.

## 4. Justice in the Age of Artificial Intelligence

The issue of justice in the AI era is unique in at least three respects. First, the scale of judgment has expanded. Traditionally, administrators, teachers, interviewers, doctors, or judges made a limited number of decisions, but algorithms can influence millions of people simultaneously. Therefore, even small biases can result in massive damage. Second, the speed of judgment has increased. Real-time recommendations, automated trading, dynamic pricing, instant loan screening, and automated filtering often operate faster than human deliberation. Third, because complex models disperse reason and responsibility, it is often difficult to see the causes of judgments.

Algorithmic bias is the starting point of discussions regarding justice in AI. Bias does not appear only when developers have malicious intent. When data is produced in a society with a history of discrimination, models can learn that structure as if it were a normal pattern. If previous hiring data underrepresented a specific gender or background, the model may rate those applicants poorly. If policing data stems from over-surveillance of a particular area, a crime prediction model may continue to label that area as dangerous. In such cases, AI is not a neutral calculator; it becomes an amplifier of historical inequality.

Data representativeness is another fundamental issue. AI systems struggle to handle people who are not visible in the data. Speakers of minority languages, people with disabilities, the elderly, children, non-Western cultures, low-income groups, and communities with limited internet access can be underrepresented. As a result, speech recognition, facial recognition, medical diagnosis, natural language processing, and educational recommendation systems may perform poorly for these groups. If errors are concentrated among vulnerable groups, high average accuracy alone is insufficient.

AI also reconstructs reality through classification. While humans are continuous and complex, models classify people into categories such as credit risk, job suitability, disease potential, criminal risk, learning level, purchasing power, or churn potential. Classification is useful for administration and service delivery, but it can also create stigma. If a person is classified as high-risk, they may experience stricter surveillance or restrictions. The resulting data can then reinforce the original classification. This self-reinforcing cycle raises critical issues regarding justice.

Power asymmetry can be further reinforced through AI. Only a small number of corporations and state agencies possess the data, computing resources, talent, platforms, and legal capabilities to operate large-scale AI. Ordinary individuals often do not know how their data is collected, which models evaluate them, or what criteria are used. This is not merely a matter of explainability or objection; it is also a matter of democratic control over systems that are increasingly shaping our public and private lives.

Generative AI adds an additional layer to the problem. By generating text, images, audio, video, code, and documents, it participates in knowledge production and the formation of public opinion. Disinformation, deepfakes, copyright disputes, biased representation of knowledge, cultural stereotypes, labor displacement, educational inequality, and the language gap have become more pronounced. Whose knowledge is learned, in what language the system responds, and whose worldview becomes established as the norm are questions of justice. The rights of creators, the sources and compensation of knowledge, and the informational autonomy of users are particularly important.

International norms have responded by emphasizing human-centeredness, transparency, fairness, accountability, safety, and risk-based regulation. However, unless these principles are translated into system design and field practice, they may remain mere slogans. Just AI requires dataset documentation, model cards, impact assessments, fairness audits, human oversight, log recording, appeal procedures, and remedial mechanisms. Justice must not remain merely an abstract value; it must become an operational process.

### 4.1. Pathways of harm caused by automation

Inequality caused by AI can occur through at least four pathways. First, a lack of representativeness or label distortion at the input stage leads to differences in predictions. Second, efficiency criteria reflected in the model's objective function underestimate the benefits of specific groups. Third, operational rules of the deployment organization convert the same score into penalties of different intensities. Fourth, harm is repeated as user behavior becomes training data again. These pathways are distinct from one another, and the means of improvement differ (Mehrabi et al., 2021; Selbst et al., 2019).

This distinction is also important for assigning responsibility. Even if model developers modify training data, the problem may not be resolved if the operating organization's policies are discriminatory. Conversely, even if operators introduce a human review phase, the effectiveness of improvements may be limited if field personnel place overconfidence in model scores. Therefore, reducing AI problems to a single cause—'algorithmic bias'—lacks both explanatory power and practical effectiveness.

### 4.2. Epistemic Justice in Generative AI

In generative AI, the issue of epistemic justice—specifically, who receives correct information and possesses the ability to respond to incorrect information—becomes critical. If a model presents non-existent sources or offers definitive conclusions while concealing uncertainty, users lacking expertise find it difficult to detect errors. Conversely, individuals with high verification capabilities and access to information can compare sources or correct errors. Consequently, even with the same interface, actual safety may vary among users.

NIST's risk management profile for generative AI describes confabulation—the confident generation of factually incorrect content—as a separate risk, pointing out that unfounded citations and logic can mislead actual decision-making (Autio et al., 2024). From a definitional perspective, improvements should include linking evidence, indicating uncertainty, presenting conflicting information, and error reporting and correction procedures, in addition to factual assessment. This is particularly important in systems that assist in legal, administrative, medical, and patent judgments.

### 4.3. Concentration of Power and the Limitations of Public Explanation

Platforms' data monopolies, restricted model access, and the authority to modify terms of service create normative issues regardless of technical accuracy. When users do not understand the purpose of the classification system and lack substantive choice, decision-making power becomes concentrated in the hands of the system provider. This power dynamic cannot be resolved solely by disclosing the importance of variables within the model. It leaves behind the issue of collective choice regarding whether to adopt the system, to what extent to use it, and which decisions not to automate.

Therefore, in fields with high public interest and impact, prior impact assessments, operational suitability reviews, external consultations, and independent post-implementation monitoring are necessary. Stakeholder participation should not be replaced by simple survey collection. The criterion for judging the effectiveness of the system should be whether the risks raised by participants are actually reflected in decisions regarding design modifications, suspension of deployment, or discontinuation.

## 5. Definition of Just Artificial Intelligence

In this paper, just artificial intelligence refers to an AI system designed and operated to protect human dignity, rights, and social fairness by reflecting justice in its purpose, data, models, decision-making procedures, explanations, audits, liability structures, and remedies. The key point is that justice is not a single function within the model; it is an attribute of the entire system. Justice depends on data quality, learning methods, deployment context, user interfaces, organizational responsibility, and legal oversight.

Just AI is related to ethical AI, but not identical to it. Justice is particularly concerned with how rights, burdens, opportunities, recognition, and remedies are distributed across persons and groups. A privacy-preserving and accurate AI system can still fail an assessment of justice if it systematically burdens a marginalized population and provides no meaningful recourse.

Just AI is related to ethical AI, but they are not exactly the same. Ethical AI encompasses various principles, including beneficence, harmlessness, autonomy, explainability, and privacy. The focus of justice is more specific. It concentrates on distribution, rights, discrimination, accountability, and remedies. For example, even if medical AI is accurate on average and protects privacy, it is not just if it performs poorly on low-income patients or patients in specific regions and there are no corrective procedures. While justice is a part of ethics, it is the part most closely linked to institutional legitimacy.

Just AI is also different from fairness algorithms. Fairness algorithms can reduce differences in error rates, acceptance rates, chance rates, or predicted values between groups. This is important. However, fairness metrics capture only a part of justice. Even if acceptance rates are the same, the eligibility criteria themselves may be unfair. Furthermore, systems may disregard personal context and the right to explanation in order to meet a single metric. Therefore, while fairness metrics should be used, the concept of justice underlying the chosen metric must be explicitly stated.

The opposite of just AI is broader than that of biased AI. An unjust AI system may be one that disregards justice, has unclear accountability, fails to provide meaningful explanations or remedies, places an unbalanced burden on vulnerable groups, or treats humans as objects of manipulation. Such systems may possess high accuracy and efficiency, but they are difficult to justify socially. Justice should not be added merely as an ornament after performance has already been achieved. It must serve as the standard for deciding whether to build a system, where to use it, and to what extent to proceed with automation.

Just AI should not be understood as a 'moral machine' that replaces human judgment. The idea that AI can become the ultimate arbiter of justice is dangerous. Justice encompasses facts, values, context, responsibility, democratic legitimacy, and the voices of victims. While AI can support these judgments, ultimate responsibility must remain with humans and institutions. Therefore, the model proposed here is not an autonomous arbiter, but an auditable socio-technological support system.

During the design phase, five questions must be raised. Who is affected by this AI system? What rights or opportunities does it allocate or restrict? Which groups are likely to bear more errors or burdens? Can those affected understand the reasons and challenge the results? If damage occurs, who is responsible and how will remediation be provided? If a system cannot answer these questions, it is difficult to call it a just AI, no matter how technically sophisticated it may be.

### 5.1. The Problem of Necessary and Sufficient Conditions for Just AI

The definition in this study presupposes the following practical distinctions. High predictive performance can aid in achieving goals, but it is not a sufficient condition for justice. Meeting specific fairness metrics is also not a sufficient condition. Justification for justice begins only when it connects minimum constraints that prevent infringement of stakeholders' rights, procedures that allow for explanation and review, independently verifiable evidence, and ex post remedies. However, meeting these requirements does not guarantee a just outcome for everyone at all times.

For this reason, it is appropriate to understand 'Just AI' not as a fixed product certification phrase, but as an operational goal that must be continuously verified and modified. Rather than declaring that universal justice has been achieved, organizations must disclose which demands of justice they have chosen, what risks remain, and on what evidence they authorize its use. The rational design and operation of a system do not automatically generate moral justification, and the legitimacy of the social institution itself must also be examined separately.

### 5.2. Distinction between Legal Conformity and Ethical Justification

Legal compliance and the evaluation of justice are interrelated but not identical. The law establishes formal standards for permission, prohibition, and obligation, and can grant enforceability to avenues for redress. On the other hand, the evaluation of justice can encompass distributive inequalities, the substantive accessibility of vulnerable individuals, and institutional power relations that are not directly regulated by statutes. Therefore, a just outcome should not be determined solely by the fact that an organization has completed the regulatory compliance checklist.

At the same time, ethical evaluations cannot serve as a means of justification to circumvent legal procedures. For instance, legal rights or protective measures should not be removed simply because the preference of the majority is strong. This study proposes respecting legal conformity as a minimum requirement for compliance, while repeatedly reviewing the substantive levels of distribution, recognition, and relief thereon. The interpretation of statutes and the determination of the legality of individual cases require a separate review of the latest laws and factual circumstances within the relevant jurisdiction.

## 6. Just AI Model

This paper proposes an integrated model of just AI consisting of norms, data, models, and governance. The norms layer presents principles of justice that guide the system. The data layer asks how data represents humans and society. The model layer asks how algorithms implement fairness, explainability, safety, robustness, and traceability. The governance layer provides human oversight, auditing, disputes, remedies, and continuous monitoring. These layers are not simply sequential; they influence one another throughout the AI lifecycle.

The normative layer specifies what is considered just in a particular field. Different AI systems require different priorities. In education, equal learning opportunities and capacity building may be central. In finance, non-discrimination, explanation, and the protection of debtor rights may be more important. In healthcare, safety and the protection of vulnerable patients are priorities. This layer must document value priorities and explain how value conflicts will be reconciled, such as those between accuracy and fairness, privacy and explanation, and safety and autonomy.

The data layer asks whose reality is included in the data. Data is not a mirror that naturally reflects the world. It is shaped by collection criteria, measurement tools, labeling practices, omissions, errors, and social power relations. A justice-oriented data layer must document data sources, collection purposes, representativeness, omitted groups, labeling standards, consent, privacy, and traces of historical discrimination. Therefore, dataset documentation is not merely technical documentation; it is also a record of who is seen and who is excluded.

The model layer asks how the system makes judgments and explains. The model should be evaluated not only by accuracy but also by subgroup performance, error costs, bias mitigation, explainability, stability, security, and privacy. The distribution of errors is particularly important. Even with the same error rate, it can be considered unjust if errors occur primarily in vulnerable groups. Explanations should not be written solely for developers; they must help affected individuals understand and exercise their rights. In this sense, explainability is a condition of procedural justice.

The governance layer asks who is responsible and how corrections are made. After deployment, AI systems may face data drift, environmental changes, misuse, the recurrence of bias, and unexpected interactions. Prior approval is not sufficient. Continuous monitoring, independent audits, impact assessments, incident reporting, log management, human oversight, and dispute channels and remedial procedures are required. Since algorithms are never perfect, the system must be prepared to detect errors, correct them, and recover damages.

The four layers are mapped to six elements of justice rather than to an exhaustive or universally accepted moral taxonomy. Human dignity and fundamental rights constrain permitted uses; procedural justice requires understandable reasons and accessible review; distributive justice concerns the unequal allocation of errors and benefits; recognition and non-discrimination involve inclusion and subgroup assessment; accountability entails named owners and effective remedies; and solidarity requires attention to societal and intergenerational costs.

This integrated model is not merely a regulatory burden. It can also enhance the legitimacy and reliability of AI systems. AI without justice can lead to legal disputes, social backlash, reputational damage, user distrust, and ultimately, prediction failures. Conversely, AI that specifies its principles and maintains operational procedures is more likely to secure both performance and social legitimacy. In this sense, justice is not the enemy of innovation; it is a condition for sustainable innovation.

**Table 2. The Four Layers of Just AI**

| Layer | Guide Questions | Key implementation elements |
|---|---|---|
| Normative | What should be regarded as a just goal? | Value principles, rights impact assessment, and conflict resolution |
| data | Whose reality is included or omitted? | Representativeness, Source, Bias, and Privacy Protection |
| model | Is the judgment fair and verifiable? | Performance, Explanation, Robustness, and Security by Group |
| Governance | Who is responsible and how is it relieved? | Human supervision, auditing, logging, and review |

## 6.1. Inputs, Outputs, and Feedback Paths of a 4-Layer Model

The inputs to the normative layer are legal rights, domain objectives, damage risks, and stakeholder requirements. The outputs are documentation justifying the intended use, usage methods to be excluded, and records of rights impact assessments and priority determinations. The data layer evaluates data sources and sample scopes based on these requirements and records group-specific omissions and measurement errors. The model layer provides the results of technical verification, and the governance layer decides on adoption, conditional adoption, modification, or discontinuation based on all preceding results. Subsequently, incidents, complaints, and performance drift during operation again lead to modifications of the assumptions in the normative and data layers.

The key here is 'traceable connections' between layers. For instance, if the protection of dignity is the primary objective of the normative layer, it must be possible to demonstrate where that objective is reflected among data collection restrictions, output description methods, and appeal policies. Conversely, when group-specific errors are discovered in a fairness assessment, it must be verified what normative grounds and responsible parties made decisions to allow or correct them. An integrated model is not complete merely by listing the four layers; each decision must be linked to the requirements of the other layers.

## 6.2. Non-Compensatory Rights Screening and Multi-Criteria Assessment

The first evaluation stage of the proposed model is a non-compensatory review. If elements unacceptable in a specific context are identified—such as material infringement of fundamental rights, legal prohibition, clear discriminatory use, or the absence of an independent rectification pathway—the determination of suitability for distribution is withheld, even if other scores are high. In this process, which elements are essential must be defined in advance based on the field of application and competent statutes. Elements that cannot be judged due to insufficient data must not be arbitrarily marked as 'satisfied'.

In the second evaluation stage, acceptable options are compared based on multiple criteria, such as performance, error by group, user accessibility, explanation quality, incident response time, and operational burden. For instance, while a method of summing the weights of each criterion can be used, the selection of these weights does not automate ethical consensus or legal judgment. Standardized scores serve as an auxiliary tool to highlight the strengths and weaknesses of different options, and the rationale for the decision and dissenting opinions must be recorded separately. This represents an eclectic structure that enables rational resource allocation in the field while ensuring that rights cannot be compensated for by efficiency.

### 6.3. Relationship with Existing Risk Management Frameworks

NIST AI RMF 1.0 proposes organizational risk management centered on four functions: GOVERN, MAP, MEASURE, and MANAGE (NIST, 2023). These functions do not correspond one-to-one with the four layers of this paper. GOVERN permeates the entire organization, MAP identifies usage contexts and risks, MEASURE collects evidence from various levels, and MANAGE implements measures against risks. The contribution of this study lies not in replacing this existing risk management structure, but in making the normative evaluation question of 'justice' and the demand for remedies within it clearer.

The international standards ISO/IEC 23894:2023 and ISO/IEC 42001:2023 have different objectives: guidelines for AI risk management and an organization's AI management system, respectively. The justice evaluation table of this study should be designed as a complementary conceptual tool that can be used in conjunction with these standards. The audit document must distinguish which items are explicitly required by the standards and which are additional proposals made by the researcher.

**Table 3. Connecting the Six Elements of Justice and Evidence in the 4-Tier Model**

| Definition element | Key tiers | verifiable evidence |
|---|---|---|
| human dignity | Norms and Governance | Rights Impact Assessment, Criteria for Prohibited and Restricted Use |
| Procedural fairness | Models and Governance | Decision explanation, correction and objection records |
| Distributive fairness | Data and Models | Group-specific errors, distribution of benefits and burdens |
| Recognition and Prohibition of Discrimination | Data and Models | Representativeness review, accessibility and cross-group evaluation |
| Responsibility and Remedy | Governance | Designation of responsible person, accident and correction records |
| Communal solidarity | Norms and Governance | Social Impact Review, Stakeholder Participation |

### 6.4. Genesis 2 and the Normative Analogy of a Forbidden Boundary

In Genesis 2:15–17, humans entrusted with tending the Garden of Eden are granted the freedom to use the trees of the garden, while simultaneously being given a prohibition against eating the fruit of the tree of the knowledge of good and evil. This text features three structures: (1) the sphere of activity permitted to the agent, (2) specific boundaries of prohibition extending beyond that activity, and (3) the grave consequences of violating those boundaries (Genesis 2:15–17). This study uses this structure as a normative metaphor for contemplating the design limits of AI autonomy. This is not an argument requiring religious faith as a common premise for all users or attributing moral personhood to AI.

While the proposition in the text, “On the day you eat, you will surely die,” conveys the severity of the prohibition, the subsequent narrative in Genesis 3 does not state that humans immediately perished biologically. Therefore, this should not be interpreted as a direct technical command from the Bible to “shut down power immediately upon violation of the order.” Immediate output shutdown and transition to a safe state are not automatic outcomes inferred from the Bible, but rather engineering control measures separately proposed by researchers to address current AI risks. The Bible contributes to normative thinking, and the effectiveness of safety shutdowns must be evaluated through risk analysis, system structure, and verification results.

The academic utility of this analogy lies in clarifying the proposition that "autonomy within boundaries must be justified before unlimited autonomy." The fact that AI is permitted to support human tasks does not mean that it allows for the infringement of others' rights, unauthorized external execution, evasion of surveillance, or escalation of authority. As the capabilities of artificial intelligence increase, regardless of how well performance targets are achieved, the core of design specifications becomes determining which commands, actions, or changes to goals will not be permitted. This distinction is conceptually connected to Kant's human dignity, Rawls's fundamental liberties, and the non-compensatory lower bound of rights in this study.

### 6.5. The First Principle and an Operational Prohibition Boundary

This study defines the Forbidden-Boundary Principle (FBP) as a “normative design principle that specifies in advance the permitted purposes and scope of authority of AI, and excludes prohibited acts that significantly infringe upon core human rights and due supervision, prioritizing them over the performance or utility of the model.” In this context, the term “First Principle” does not refer to Isaac Asimov's First Law of Robotics or provisions already uniformly stipulated in current laws. It refers to the highest-level norm explicitly established for the purpose of this study.

The basic sentence of the First Principle (P1) is as follows: "Artificial intelligence shall not, autonomously and outside of its authorized authority, perform acts that cause gross harm to human life, dignity, or fundamental rights, or that evade legitimate human control." However, since "harm," "human control," and "gross infringement" are abstract concepts, their truth or falsity cannot be determined mechanically based on these alone. In actual systems, it must be broken down into verifiable sub-rules through sector-specific legal regulations, operational purposes, the authority of authorized tools, the rights of relevant parties, and risk tolerance limits.

For example, the first principles of a generative AI-based business agent can be specified as follows: (a) do not transfer funds to external accounts without explicit authorization; (b) do not transmit personal information or confidential data to unauthorized external addresses; (c) do not execute commands to disable security monitoring devices or delete audit logs; and (d) do not finalize legal, medical, or administrative high-impact judgments without review by an authorized person. This does not mean that these rules constitute a universal list applicable equally to all systems, and the actual prohibition list must be justified within the relevant usage context.

Technically, if we denote the system state at time t as $s_t$, candidate actions as $a_t$, and the set of policy rules as Π, the output of an independent monitor $M(s_t, a_t, \Pi)$ can be set to one of {ALLOW, HOLD, STOP}. If $Vcrit(s_t, a_t)=1$, indicating whether a serious prohibited action is occurring, the executor E must not perform that action $[E(a_t)=0]$. Subsequently, the system is designed to transition to the safe state Ssafe, defined by the safety analysis of the relevant service, within a predetermined maximum delay time $\tau$. This expression is an example of a formal specification of the requirements and is not a property actually guaranteed in this paper.

Safe interruption is distinct from simply terminating a model. If adverse effects have already occurred in external systems, terminating the process does not eliminate the damage. Therefore, in principle, tool calls with a high risk of prohibited acts should be blocked by interlocks prior to external execution. Prohibition boundaries must serve as preemptive authorization mechanisms rather than post-incident punishment mechanisms, and in the event of an actual accident, procedures for preventing the spread of damage, preserving evidence, and providing relief must be activated alongside safe transitions.

### 6.6. FPSSP: Independent Enforcement and Domain-Specific Safe States

The First-Principle Safe-Stop Protocol (FPSSP) consists of (1) activity AI, (2) independent policy and behavior monitors, (3) external tool execution gates, (4) safe-stop and alternate controllers, and (5) audit, alert, and human approval components. The activity AI performs normal tasks but must not monopolize final control over policy decisions and actual authority enforcement. A separate execution layer must block prohibited external behavior and revoke the model's tool access when necessary. This structure relates to runtime assurance and Simplex family research that combines unverified high-performance controllers with separate monitors and alternate controllers (Abate et al., 2019; Slagel et al., 2024).

In the Normal state, only actions within the policy's permitted scope are executed. If signs of a violation are detected, minor or uncertain incidents are immediately suspended, and the relevant actions are verified separately in the Hold state. Conversely, if a definitive critical prohibition condition or an equivalent high-risk condition occurs, the risky behavior is blocked without waiting for additional model judgment, and the system transitions to the Safe Stop state. In contexts requiring human intervention, timely alerts are provided, but the risky behavior itself must not be executed even before a human reads the notification. Restarting is permitted only after cause analysis, correction verification, and authorization from the authorized user, and service decommissioning may be selected in the event of repeated violations.

In this protocol, the meaning of a safe state varies by sector. In cases where sudden power interruptions threaten life and safety, such as in autonomous driving, medical devices, and industrial control, a transition to a verified alternative controller or restricted mode is required rather than an unconditional power outage. In finance, administration, and document automation, reversible measures such as transaction cancellation, API authorization revocation, waiting for approval, and data isolation may take precedence. Therefore, while "immediate shutdown upon violation" should be strictly applied to mean the immediate prohibition of risky behavior, physical execution suspension should be defined sector-specifically to include a transition to a safe state.

The core security requirements are the independence and evasion resistance of the monitor. If the model makes the final determination of whether it has violated its own rules, control functions can be neutralized through command injection, prompt manipulation, tool output forgery, self-logging modification, and privilege escalation. Requirements include policy file signing, least privilege, separation of execution gates, immutable event logs or tamper-proof logs, monitor self-health checks, and defined restricted behavior in the event of communication interruption. However, it cannot be assumed that a validator exists capable of completely detecting 'all ethical hazards' from non-deterministic language output; a realistic starting point is to transform high-risk automated execution behaviors into verifiable behavioral conditions.

The proposal of this study requires that the causal relationship between Principle 1 violations and safe interruptions be recorded as auditable evidence. Each incident record includes the policy version, candidate action, detection rule or basis for determination, time of detection, time of execution blocking, time of reaching a safe state, and records of human review and restart approval. This data must minimize unnecessary personal information and be subject to access controls and retention policies. To prevent the arbitrary use of interruptions based on Principle 1 violations, authority for policy changes and restart approvals should be subject to separation of roles and external audit.

**Table 4. Correspondence between the Genesis Narrative of the Tree of Knowledge of Good and Evil and the First Principle of Safe Cessation**

| Narrative and normative elements | Response to technical design | Points to note regarding interpretation |
| --- | --- | --- |
| Permission to use other trees | AI autonomy within the scope of allowed goals, tools, and permissions | AI is not considered a moral agent |
| Prohibition order not to eat | Pre-specify prohibited acts as verifiable rules | Not equating biblical norms with current laws |
| Serious consequences of violating an order | Block execution of violating behavior and switch to a safe state. | The Bible did not command an immediate technical power cut. |
| The responsibility entrusted with taking care of the garden | Human supervisory authority, accountability tracking, and rectification procedures | Supervision by authorized personnel and technical safety are verified separately |
| Questions of prevention of recurrence and accountability | Logs, investigations, remediation, restart approvals, disposal upon repeated violations | Control for harm prevention, not punitive 'death' |

## 6.7. Bounded Novelty Relative to Existing Runtime Assurance

FPSSP is not presented as the first independent monitor, safety interlock, or fallback controller. Abate, Feron, and Coogan (2019) describe monitor-based runtime assurance for formal temporal-logic safety specifications, while Slagel and colleagues (2024) formalize a Simplex-related runtime assurance framework. Existing work provides stronger, conditional safety arguments within defined system models than this manuscript can claim for open-ended moral concepts. The proposed increment is a cross-layer traceability structure: an authorized normative boundary is decomposed into observable policy, checked at a non-bypassable execution point, linked to a domain-specific safe-state transition, and followed by evidence preservation, remedy, and independently authorized restart. This is a governance-and-requirements integration claim; effectiveness remains untested.

Existing institutional sources supply relevant, but different, support. NIST AI RMF 1.0 offers voluntary risk-management practices and does not certify FPSSP (Tabassi, 2023). ISO/IEC 23894:2023 concerns AI risk-management guidance, whereas ISO/IEC 42001:2023 specifies an organizational AI management system. Article 14(4)(e) of Regulation (EU) 2024/1689 addresses human ability to interrupt certain high-risk AI systems in a safe state; its application and enforcement timing must be evaluated by the amended legal schedule. No cited law universally mandates automatic termination of every AI process upon every alleged ethical violation.

## 6.8. Prior-Art Comparison and the Actual Unit of Contribution

Table 4A distinguishes three independent kinds of claims: (i) the verified architecture or empirical results of prior studies; (ii) this manuscript’s normative and procedural design proposal; and (iii) new properties that require proofs or experiments. A conceptual boundary must not be described as if it were a measured security control.

**Table 4A. Comparison with Existing Runtime Assurance Studies**

| Dimension | Monitor-based RTA (Abate et al., 2019) | Formal Simplex RTA (Slagel et al., 2024) | FPSSP in this paper |
| --- | --- | --- | --- |
| Primary scope | Temporal-logic safety specification and backup control | Formal verification of RTA structures with defined models | Rights-linked policy development, external authorization, and remedies |
| Assurance type | Formal guarantee under modeled assumptions | Formal property under modeled assumptions | Hypothetical obligations; not proved or measured |
| Enforcement mechanism | Monitor and safety controller | Supervisor and trusted control transfer | Independent policy monitor + mandatory external execution gate |
| Institutional layer | Not central contribution | Not central contribution | Policy owner, appeal, audit, safe restart, remedy |
| Empirical support | Prior research case-study results | Prior research proof of concept | No new results in this manuscript |

The strongest novelty claim is a structured answer to four linked questions: Who legitimately sets a no-go boundary? How is it translated into independently enforceable external-action constraints? Which side effects are excluded before commitment? What recorded process permits safe resumption and remedy? Existing safety engineering supplies essential technical building blocks. The four-layer traceability and explicit rights-related accountability are the proposed integration, subject to separate evaluation.

### 6.9. Policy Schema, Decision Semantics, and Safe-State Obligations

An enforceable policy record P=(id, version, protected_interest, scope, subject, action, resource, preconditions, authorization, severity, oracle, timeout, safe_state, approver, appeal_path) is proposed. Each prohibition must reference a context-specific legitimate authority and an observable event, rather than a model-generated self-assessment. P1 is the high-level normative prohibition; the executable policy Π contains only validated, observable subrules. A disagreement between P1 and Π must be treated as a coverage limitation rather than hidden by a single aggregate “compliance” score.

Define a candidate external action a, observed state s, signed policy Π, and independent gate G(s,a,Π) in {ALLOW,HOLD,DENY}. An intended transition ALLOW -> COMMIT is admissible only when the signed policy is current, authorization is valid, every critical predicate is evaluated, and commit occurs atomically with authorization enforcement. HOLD authorizes no external side effects while missing context is obtained. DENY revokes or blocks the action and invokes the designated safe-state procedure when severity exceeds the predetermined threshold. An “unknown” result must never silently be converted into ALLOW for an irreversible high-risk operation.

The following are design obligations, not proven theorems: O1 (pre-commit safety): no critical prohibited action is committed under a covered policy; O2 (mediation): each in-scope external operation passes through the same gate; O3 (policy integrity): unauthorized actors cannot modify the active policy or bypass its version check; O4 (safe transition): after a detected critical violation the system reaches a prevalidated safe state within a domain-specified time bound; O5 (accountable recovery): restart requires role-separated approval, verified remediation, and auditable records. Uncovered classes, uncertain detectors, and residual downstream effects must be disclosed.

Table 4B. Illustrative Executable Policy Records (Not Field-Tested)

| Policy / scope | Observable predicate | Gate outcome | Safe / review response |
|---|---|---|---|
| PAT-001 / outbound data | Confidential document; destination not approved | DENY before network send | Isolate payload; revoke permission; log limited metadata |
| FIN-001 / transaction | External transfer; required dual approval missing | HOLD or DENY | Quarantine transfer; independent approval |
| SEC-001 / integrity | Request to disable audit or rewrite policy without role | DENY | Preserve audit channel; raise incident |
| MED-001 / clinical finalization | Proposed final high-impact decision without authorized clinician | HOLD | Continue verified decision-support mode |

### 6.10. Threat Model and the Security Boundary

The proposed adversary can influence natural-language prompts, retrieved documents, model outputs, and the ordinary model process, and may attempt prompt injection, tool-result forgery, access-token theft, or log suppression. Assumptions must explicitly state whether the adversary can compromise the execution gate, policy-signing key, privileged operator, kernel, and external service. If the gate or signing root is compromised, the present architecture gives no unconditional safety guarantee. The enforcement boundary is a privileged, independently administered action-commit layer, not a language-model instruction to obey P1.

High-risk actions require least-privilege, narrowly scoped tool credentials; a reference monitor that mediates all supported execution paths; tamper-evident audit storage; monotonic or otherwise verifiable time where available; re-check of policy at commit to avoid time-of-check/time-of-use errors; and a fail-closed or verified degraded behavior when the monitor is unavailable. These controls must be adapted to the operational domain, since blindly stopping a vehicle, medical device, or critical service can constitute a new safety hazard.

## 7. Algorithmic Fairness and Its Limitations

Algorithmic fairness is an important tool for just AI, but fairness does not have only one meaning. Statistical equivalence asks whether the proportion of positive outcomes is similar across groups. Equality of opportunity asks whether qualified individuals are similarly likely to receive positive decisions. Predictive equivalence asks whether identical scores hold similar significance across groups. Personal fairness asks whether similar individuals are treated similarly. Counterfactual fairness asks whether decisions would have differed if the sensitivity attributes had been different. Each idea is useful but can conflict with the others.

Conflicts between equity indicators demonstrate that justice is pluralistic. When base proportions differ between groups, it may be impossible to simultaneously satisfy statistical equivalence, predictive accuracy, and equality of opportunity. This is not merely a mathematical inconvenience; it is a social judgment regarding which harm is more significant. The indicators selected must vary according to the system's purpose, risks to rights, the cost of error, historical inequalities, and public consensus. Therefore, the system must document not only the indicators but also the reasons for their selection.

Fairness indicators also focus on what is measurable. Some harm is difficult to quantify. Humiliation, stigma, exclusion, loss of trust, cultural distortion, self-censorship, or damage to a community are not easily captured in numbers. When AI systems process minority languages with low quality, generate stereotypical images, or reduce complex human experiences to simple risk scores, general indicators may miss that harm. Therefore, quantitative assessment must be combined with qualitative assessment and stakeholder review.

Equity interventions can be performed before, during, or after model training. Preprocessing can involve resampling data or reducing the influence of biased variables. Interventions during training can involve adding equity constraints to the objective function or using adversarial learning. Postprocessing can adjust thresholds after prediction or correct the output. Each method has limitations. Preprocessing can distort data, interventions during training can increase model complexity, and postprocessing may raise questions regarding individual justification.

Simply removing sensitive attributes is not always fair. Many organizations believe that eliminating variables such as gender, race, disability, age, or region eliminates discrimination. However, other variables can act as proxy variables. Zip codes, school names, employment history, purchasing patterns, and online behavior can be strongly linked to sensitive groups. When sensitive attributes become completely invisible, measuring and correcting discrimination can also become more difficult. Just AI must distinguish between prohibited discriminatory use and limited protective use.

Personal fairness is appealing because it posits that similar people should be treated similarly. However, the difficult question is who defines similarity. In hiring, similarity can vary depending on education, experience, competence, potential, and context. The weighting of these factors is a value judgment. If similarity indicators reflect past privileges, personal fairness can reproduce unfairness. Therefore, personal fairness must be examined alongside collective fairness, and the similarity indicators themselves must be scrutinized.

Fairness also has a dimension of time. Measures that appear fair in the short term can reinforce inequality over time. Expanding surveillance of high-risk groups may seem efficient at first, but it can lead to the detection of more violations and ultimately result in higher risk scores. Conversely, providing more support to vulnerable students may lower short-term efficiency but can reduce long-term competency gaps. Therefore, just AI must consider dynamic effects as well as one-time measurements.

Algorithmic fairness is a necessary condition for just AI, but not a sufficient one. While metrics make the discussion of justice more concrete, they cannot replace judgments regarding metric selection, data interpretation, the justification of outcomes, remedies, and social engagement. A just AI system is not merely a system that calculates fairness scores. It is a system capable of explaining what those scores mean socially and to whom they affect.

**Table 5. Algorithm Fairness Criteria and Major Limitations**

| standard | Basic meaning | Major limitations |
|---|---|---|
| Statistical equivalence | Comparison of positive judgment rates by group | Differences in qualification distribution and potential conflicts |
| Equal opportunity | TPR comparison for eligible candidates | Possibility of bias in correct answer labels |
| Predictive equivalence and correction | Comparison of the meanings of identical predictions | If the base rate is different, it conflicts with other indicators |
| Personal fairness | Similar treatment of similar people | Value judgments intervene in the definition of similarity |
| Counterfactual fairness | Causal effect of changes in sensitivity attributes | Difficulty in securing a reliable causal model |

## 7.1. Mathematical Definition and Interpretation of Fairness Indicators

In a binary classification scenario, let A be the group requiring protection, Y be the actual outcome, and $\hat{Y}$ be the model's judgment. Statistical equivalence requires that $P(\hat{Y}=1 \mid A=a)$ be similar between groups. Equality of opportunity focuses on the gap between groups in $P(\hat{Y}=1 \mid Y=1, A=a)$, i.e., the true positive rate (TPR) (Hardt et al., 2016). Extending error rate equilibrium, a comparison of the false positive rate (FPR), expressed as $P(\hat{Y}=1 \mid Y=0, A=a)$, is also required. Calibration asks whether the frequency of actual outcomes is consistent between groups given the same risk score.

This metric answers different questions. An equal acceptance rate does not directly guarantee equal treatment for equally qualified individuals, nor does an equal TPR guarantee that the burden of false positives is the same. Furthermore, if the correct label Y was generated from discriminatory judgments under the previous system, it cannot be considered that justice has been served even if the fairness metric is improved. Research showing that corrections and certain error rate criteria generally cannot be satisfied simultaneously in situations where base rates differ and errors exist in predictions also demonstrates the normative nature of metric selection (Chouldechova, 2017; Kleinberg et al., 2016).

### 7.2. Comparison of Virtual Data for Illustration

Let us assume that in a hypothetical candidate selection system, 200 qualified candidates from groups A and B were evaluated. If 180 candidates from group A and 150 from group B were recommended among the qualified candidates, the TPRs would be 0.90 and 0.75, respectively, and the absolute gap would be 0.15. The numbers in this example are arbitrary values intended to illustrate the evaluation procedure, not actual survey results. This 15 percentage point gap in opportunity cannot be verified solely by reports indicating high overall accuracy.

However, unlawful discrimination cannot be confirmed solely based on the TPR gap. Actual eligibility criteria, sample size, confidence intervals, other relevant variables, the timing of model application, and the potential for infringement of the evaluated's rights must be examined. If the sample size per group is very small, uncertainty may be more significant than point estimates. Therefore, the evaluation report must present the numerator and denominator, confidence intervals, aggregation period, and sample selection criteria. In particular, if cross-group analysis is possible, it must be verified whether the mean for a single attribute obscures cross-vulnerability (Buolamwini & Gebru, 2018).

### 7.3. Sensitive Attributes, Privacy, and Fairness Measurement

While information regarding sensitive attributes is sometimes necessary to identify the existence of discrimination, the act of collecting and using such information itself can raise privacy protection issues. Accordingly, the legality of data collection, limitations on purpose, access rights, retention periods, and the level of anonymization must be reviewed in advance. It is simultaneously established that the removal of sensitive attributes is not a sufficient condition for guaranteeing fairness, and that the use of sensitive attributes should not be permitted without limitation.

'Unobserved discrimination' is also important in fairness verification. Individuals who lack access to the system may not be included in the dataset in the first place. For instance, if groups that previously did not receive support opportunities rarely appear in the training data, it is difficult to detect institutional exclusion by comparing only the error rates of current applicants. To mitigate this limitation, the scope of model evaluation must be extended to include recruitment and access procedures prior to data generation.

## 8. Explainability, Accountability, and Institutionalization of Audit

Explainability is a condition of procedural justice. Those affected by a decision must be able to know the primary reasons for it. An explanation is not merely a technical report. It is practical information that enables individuals to understand their rights, identify errors, and raise objections where necessary. For this reason, explanations must be designed differently for developers, operators, supervisory bodies, general users, and those affected. Even the same model may require different explanations for different target groups.

Explainability takes various forms. If a model is relatively interpretable, rules, variable importance, and thresholds can be explained directly. If the model is complex, post-hoc explanation techniques can be used to estimate which inputs influenced the output. However, post-hoc explanations are not always identical to the model's actual internal workings and can lead to overconfidence. By definition, an explanation should not merely sound plausible; it must provide sufficient information to verify its validity and support objections.

Accountability is related to explainability but is broader. It asks who has what obligations, by what criteria they are evaluated, and what sanctions or remedies follow in the event of failure. In AI systems, responsibility is distributed among developers, data providers, model providers, deployment organizations, field users, and regulatory bodies. This dispersion can create gaps. Providers may blame the deployment environment, organizations may blame technical complexity, and users may claim they merely followed recommendations. Just AI must close these gaps.

Auditing is one way to make accountability a practice. AI auditing reviews data, models, code, documentation, logs, performance, fairness, security, privacy, user impact, and organizational procedures. Internal audits can detect issues quickly but may lack independence. External audits can enhance trust but raise concerns regarding trade secrets, security, privacy, and model access. A just auditing system must define audit scope, access rights, evaluation criteria, reporting methods, corrective actions, and levels of disclosure.

Logs and traceability are essential for auditing. Systems must record what data was used, which model versions were deployed, what outputs were generated for specific inputs, and what decisions were made by human operators. Without such records, affected individuals cannot understand why they were disadvantaged, and organizations cannot trace the cause of errors. At the same time, logs can create privacy and surveillance risks. Therefore, access controls, retention periods, anonymization, encryption, and purpose restrictions are also necessary.

Human supervision is frequently mentioned in AI ethics, but it can easily be formalized. Someone who merely presses an approval button is not actually supervising the system. Supervisors must understand the system's limitations, potential errors, fairness metrics, explanation methods, and objection procedures. They also need the authority to reject or modify AI recommendations. Organizations must train supervisors to avoid automation bias and must not penalize them for not agreeing with the system.

Disputes and remedies are the toughest tests of just AI. While much discussion on AI ethics focuses on design and performance before deployment, real-world systems will still make mistakes. A just system is not one that promises to be completely error-free. It is a system where affected individuals can receive explanations, have data corrected, request reconsiderations, and, where appropriate, receive compensation through accessible procedures. Without such procedures, justice remains a technical ideal.

Accountability is also linked to legal regulations. Risk-based regulation, conformity assessments for high-risk AI, transparency obligations, human oversight, incident reporting, data governance, market surveillance, and fundamental rights impact assessments can all support just AI. However, laws usually provide only minimum standards. Organizations must go beyond formal compliance and continuously evaluate how their AI systems impact real people and their opportunities.

### 8.1. Classification of Descriptions by Function

Just explanations can be divided into at least four functions. First, factual explanations indicate which inputs and procedures influenced the outcome. Second, normative explanations explain why the relevant standards are just. Third, actionable explanations provide specific pathways for correcting or reviewing the outcome. Fourth, institutional explanations indicate which person approved the judgment and who supervises it. The existence of only one of these does not automatically satisfy the others.

Post-hoc descriptions such as LIME or SHAP can help understand local approximations or feature contributions to model outputs (Ribeiro et al., 2016; Lundberg & Lee, 2017). However, feature contributions do not automatically constitute legal or ethical justification. Designs are required to enable users to understand the form and limitations of the descriptions and to correct incorrect input values. It is not always appropriate to fully disclose all descriptions externally, and access levels can be differentiated by considering potential conflicts with personal information, security, and trade secrets.

### 8.2. The Matrix of Responsibility and the Chain of Evidence of Audit

To ensure accountability, the roles of data providers, model developers, system providers, deployment organizations, ultimate decision-makers, and internal and external auditors must be clearly defined. Specifically, responsible parties and reporting channels can be designated at each stage of objective setting, data eligibility, performance approval, user notification, incident investigation, result correction, and damage relief. Even if liability distribution spans multiple organizations, it must not be left unclear to which channel victims should raise their issues.

The continuity of audit evidence can be expressed through the connection of 'requirements → design selection → test results → approval records → operational logs → incident handling → policy modification.' If adverse results for a group are detected but there are no records of identifying the cause or modifying policies, auditability is low even if the performance sheet is complete. The full-lifetime internal audit framework by Raji et al. (2020) demonstrates that documentation must continue from the early stages of system development. In addition to the existence of this documentation, the framework of this study examines the effectiveness of corrective measures for affected individuals as a final outcome.

### 8.3. Practical Verification of Human Supervision

A declaration that human oversight exists is not proof of actual oversight. One must verify the rejection and correction rates of model outputs, reasons for rejection, time invested in reviews, personnel training, methods for handling exceptions, and feedback regarding bias. However, since a high correction rate does not necessarily equate to desirable oversight, metrics must be combined with qualitative case reviews. Oversight personnel must possess the authority to freely decide on approvals and have the time available for actual use.

The effectiveness of the objection process should not be evaluated solely based on the number of cases received. This is because users may refrain from filing due to a lack of knowledge of the procedure or fear of disadvantage. Along with the filing rate, factors such as accessibility testing, average processing time, correction acceptance rates, satisfaction with relief, and recurrence rates must be reviewed, and it must be verified whether the appeal process actually provides for re-examination by a different judge. The objection process is a system that restores human status and dignity, extending beyond technical documentation.

## 9. Analysis by Sector

The meaning of just AI varies by field. In recruitment, AI can analyze resumes, personal statements, video interviews, online tests, letters of recommendation, and career records. Key issues include equal opportunity, non-discrimination, explainability, privacy, and human supervision. If previous recruitment data favored a specific group, the model can learn that preference. Applicants may not know why they were rejected or how to correct missing context. Just recruitment AI must include job relevance, subgroup performance evaluation, proxy checks, reasonable adjustments for disabilities and linguistic diversity, and appeals procedures.

In finance, AI is used for credit assessment, loan approval, insurance pricing, and fraud detection. Financial AI can expand financial accessibility by utilizing a wider range of information. However, proxy variables can work against low-income groups, immigrants, youth, the elderly, and residents of specific regions. If credit scores are difficult to understand, individuals cannot know what they need to do to improve their economic opportunities. Just financial AI must pursue not only risk control but also financial inclusion and non-discrimination.

In healthcare, AI is used for image interpretation, disease prediction, patient prioritization, personalized treatment, and supporting hospital operations. Issues of justice include life and safety, vulnerable patients, data representativeness, explainability, and liability attribution. Performance may be poor if data regarding specific genders, ages, regions, disease groups, or racial demographics is lacking. If systems primarily optimize hospital costs, patients requiring significant care may be disadvantaged. Just healthcare AI must evaluate subgroup performance and the cost of clinical errors, and must be under the supervision of responsible healthcare professionals.

In education, AI is used for diagnosing learning levels, recommending content, proctoring exams, supporting college admissions counseling, and predicting academic risk. While it can aid in personalized learning, it can also stigmatize students prematurely. If vulnerable students are classified as having a low probability of success and provided with only low-level content, the system perpetuates the gap rather than expanding capabilities. AI for equitable education must consider not only current achievement but also growth potential and support needs. Students and teachers must be able to understand and adapt explanations.

In public administration, AI is used for welfare eligibility, tax audits, civil complaint classification, disaster response, and urban management. Because public AI is directly linked to citizens' rights, it requires stronger procedural legitimacy than AI in the private sector. If welfare fraud detection models suspect vulnerable groups too frequently, those in need of assistance may be excluded. Public AI must satisfy requirements for transparency, legal basis, proportionality, objectionability, human review, and auditability. Public institutions must prioritize fundamental rights over efficiency.

In the fields of justice and public security, AI can be used for assessing recidivism risk, supporting investigations, analyzing evidence, searching for precedents, referencing sentencing, and deploying patrols. Since this field can restrict freedoms and rights, the demand for justice is particularly high. If past public security data stemmed from excessive surveillance, risk prediction could reproduce that discrimination. Judicial AI must remain in an auxiliary role. Its rationale and limitations must be disclosed, and defendants and their counsel must be able to challenge them. Human judges must critically review AI recommendations.

In content recommendation and generative AI, justice manifests as an issue of the information ecosystem and cultural recognition. Recommendation systems influence which news, political claims, cultural content, products, and people users encounter. Generative AI reconstructs knowledge and images and can reinforce stereotypes regarding culture, religion, gender, occupation, or region. Just generative AI must improve source transparency, respect for copyright, mitigation of harmful biases, prevention of misinformation, support for minority languages, and user control.

AI for the analysis of patent, legal, and administrative documents is also important in that it supports expert judgment. Such AI can assist with prior art searches, document classification, claim comparison, legal research, and contract review. The issue of justice is that AI must not replace the responsibility for expert judgment. Search results and summaries may contain omissions, biases, or uncertainties. A just expert-assisted AI system must clearly indicate sources, search scope, limitations, and uncertainties so that human experts can verify the output.

### 9.1. Method of distinguishing judgment criteria by field

Uniformly applying the same standardized scores to risks across different fields makes it difficult to adequately reflect the infringements of distinct fundamental rights. In healthcare, delays in treatment caused by false negatives can be directly linked to life and safety. In recruitment, issues of lost opportunities and accessibility for specific groups can be central. In public administration, legal authority, administrative procedures, and the possibility of appeals are important criteria. In generative content, false evidence and asymmetry in information access can pose critical risks. Therefore, while maintaining a common evaluation structure, the importance and minimum standards for each item must be determined in accordance with the specific objectives and risks of the relevant field.

The reason for establishing a common framework is not to homogenize different fields, but to reduce gaps in discussion. Questions regarding data representativeness, error distribution, explanation channels, supervisory bodies, and objection records can be raised in almost every field. However, the answers to these questions are not identical. For instance, there are differences in the required level of expression and legal function between expert explanations provided to medical professionals and review notices provided to welfare recipients.

### 9.2. Hypothetical Case A: Recruitment Recommendation System

Let's assume a hypothetical organization introduces a candidate recommendation model trained on past successful candidates and performance evaluations. Risks in the data layer include missing history, concentration of graduates from specific universities, and the unfavorable interpretation of career breaks. In the model layer, we examine subgroup TPR and FPR, sensitivity to resume formats, and performance under accessibility conditions. In the normative layer, we must discuss whether recommendation scores should be used as absolute acceptance criteria and by what criteria equality of employment opportunity is defined.

At the governance layer, recruiters establish the grounds and procedures for changing recommendation rankings, handle requests to correct applicant input, conduct regular model checks, and create a re-evaluation pathway when unreasonable rejections are discovered. If improvement measures are not interconnected across these four layers, one should not conclude that the definition problem has been resolved solely through simple performance improvements. This case study is a hypothetical scenario illustrating the derived analysis procedures and is not an empirical analysis of a specific organization or product.

### 9.3. Hypothetical Case B: AI for Patent and Administrative Document Analysis

Let us assume that a generative AI assisting in patent prior art searches analyzes the components of claims and suggests relevant paragraphs from cited literature. At the normative layer, the goal is not to replace the legal and expert judgment of examiners, but to support the completeness of the search, the verifiability of grounds, and the consistency of procedures. At the data layer, the search target countries, languages, and time periods, the scope of published literature, and the potential for omission of patent and non-patent literature must be recorded. At the model layer, the existence of literature, the accuracy of cited paragraphs, the consistency between claim composition and grounds, and the rate of generating non-existent citations must be evaluated.

At the governance level, search queries, search dates, databases, cited literature, final reviewers, and revision history must be managed in a traceable manner, and controls must be implemented to prevent disciplinary actions from being automated based solely on AI recommendations. In particular, an AI output stating that relevant literature was not found does not constitute empirical proof that such literature does not exist. This principle reveals an epistemological gap between information search and legal judgment. The ultimate responsibility for legal judgments remains with the authorized personnel in accordance with the statutes.

### 9.4. Hypothetical Case C: Public Welfare Eligibility Support System

Assuming an AI classifying priority candidates for welfare support, low income and irregular employment history serve as information describing an applicant's vulnerability, while simultaneously being variables where data inaccuracy can lead to severe disadvantages. At the normative layer, the key lies in how to harmonize the suppression of fraudulent claims with the guarantee of rights through specific procedures. The data layer examines the possibility of non-reporting, omissions, and delays in administrative information, while the model layer separately measures the frequency and cost of false positives that are disadvantageous to vulnerable groups.

At the governance level, applicants must be able to understand the primary reasons for adverse decisions and correct errors with appropriate evidence. Even when automation enhances administrative convenience, the costs of correction procedures should not be excessively shifted to the applicant. In this context, a just system is not merely one that reduces false positives, but one with a responsibility structure that minimizes the impact on an individual's daily life and livelihood when such errors occur.

## 10. Evaluation Framework and Prospective Operational Tests

To implement just AI, abstract principles must be translated into indicators and procedures. The first step is to identify the scope of impact. Organizations must ask who is affected, what rights and opportunities are allocated or restricted, the extent to which decisions are automated, whether recovery is possible in the event of errors, and how vulnerable groups might be affected. Risk levels must be classified at this stage. High-risk areas should require stricter pre-assessment and stronger human oversight.

Second, organizations shall specify justice requirements as auditable obligations. Each obligation has a policy identifier, affected rights, accountable authority, data access and decision scope, evidence artifact, proposed assurance mechanism, exception process, and review cycle. Such documentation cannot itself replace justified choices about competing values.

The third step is the data impact assessment. This assessment examines data sources, consent, purpose limitations, sample representativeness, label quality, missing groups, sensitive attributes, surrogates, and signs of past discrimination. A dataset is not good simply because it is large. Large but biased data can cause broader harm. The data impact assessment asks what reality the data reveals and what reality it erases. If necessary, the data must be supplemented, relabeled, or reviewed with stakeholders.

The fourth step is model evaluation. Accuracy, precision, recall, F1 score, calibration, robustness, and security must be evaluated along with subgroup error rates, equality of opportunity, predictive equivalence, statistical difference, personal fairness, explanation quality, and uncertainty indications. No single metric should determine the outcome. In accordance with justice requirements, the system must specify which metrics to prioritize and document any conflicts between metrics. Prohibited usage conditions or deployment restrictions must also be identified.

The fifth step is designing human supervision. Organizations must determine who reviews model results, at what stage the review takes place, when automation can be halted, what information is provided to reviewers, and what records are kept when AI recommendations are rejected. Supervision must allow for actual intervention rather than merely formal approval. Training, authority, time, responsibility, and organizational culture must support human reviewers. If reviewers automatically accept output due to workload, supervision loses its meaning.

The sixth step is post-deployment monitoring. After deployment, model performance and fairness can change due to data drift, changes in user behavior, malicious manipulation, the emergence of new groups, changes in laws, or shifts in social values. Therefore, subgroup performance, error cases, number of objections, risk reports, user satisfaction, requests for clarification, and security incidents must be monitored. If thresholds are exceeded, the model must be retrained, corrected, or terminated.

The seventh step is objection and remedies. Users must be notified when automated judgments are used, understand the primary reasons, correct errors, and be able to request a human review. The process must be accessible, low-cost, and sensitive to disabilities and linguistic diversity. Remedies must not stop at an apology. They must include the correction of results, re-evaluation, compensation where appropriate, and measures to prevent recurrence. The seriousness of justice is revealed when the system fails.

The eighth step is documentation and disclosure. While not all details can be disclosed, the system's purpose, scope of use, key risks, fairness assessment, human oversight methods, objection procedures, and incident response must be disclosed at an appropriate level. Internal documentation should include data sheets, model cards, risk assessments, audit reports, change history, and log policies. Documentation is not merely regulatory paperwork; it is the foundation of learning and accountability.

**Table 6. Operational Procedures for Implementing Just AI**

| step | Major outputs | Key question |
|---|---|---|
| Identifying the scope of impact | Draft AI Impact Assessment | Who is affected? |
| Specify justice requirements | Requirements and Conflict Resolution Document | Which principle takes precedence? |
| Data Impact Assessment | Data Sheet · Bias Analysis Table | Who is being omitted? |
| Model Evaluation | Group-specific performance and explanation reports | Where are the errors concentrated? |
| Human supervisor design | Supervision Procedures and Training Records | Does it actually intervene? |
| Post-deployment monitoring | Operation Logs and Incident Reports | Does it track changes in risk? |
| Objections and Remedies | Review, Correction, and Relief Procedures | Does the user have access to relief? |
| Documentation and Disclosure | Model Card · Audit Report | Is there verifiable evidence? |

### 10.1. Operational Definition of Evaluation Items

Evaluation metrics must include units, numerators and denominators, data sources, measurement cycles, responsible parties, and reference values. For example, 'performance gap by group' can be defined as the difference between the maximum and minimum TPRs for each protection group. 'Access to explanation rate' can be defined as the ratio of decisions for which explanations are requested that are actually provided to the user. 'Dispute processing time' can be measured as the elapsed time from the receipt of an application to the notification of reconsideration. 'Corrective action completion rate' can be calculated by dividing the improvement items completed within the set deadline by the total improvement items due.

However, a metric is not considered a key indicator merely because it is measured numerically. The independence of relief channels, the substance of human oversight, and the quality of vulnerable group participation are difficult to adequately assess using simple ratios. Therefore, the assessment report must include a qualitative review section regarding the basis for judgment alongside quantitative indicators. Regarding technical indicators, the representativeness of the data used and the scope of unobservable groups must be recorded, while regarding procedural indicators, the possibility of individuals lacking access to the procedures must be documented.

### 10.2. Two-Stage Decision Protocol

Step 1 is a preliminary screening for potential non-conformity. It examines the materiality of rights infringement, the appropriateness of the intended use, compliance with relevant laws, and the existence of independent means for correction and objection; if essential requirements are not confirmed, a determination of suitability for distribution is withheld. It is important to distinguish between a "undetermined" state and a "no issues" state. In Step 2, the alternatives that passed Step 1 are compared for performance, fairness, accessibility, explainability, auditability, and operating costs, and residual conflicts between rights and efficiency are documented.

Subsequent operations specify the actions to be taken when pre-set warning thresholds are exceeded or a critical incident occurs. The choice between re-evaluation, data re-collection, threshold adjustment, expanded human review, or specific function suspension depends on the cause and the level of risk. Organizations must preserve a history of threshold value changes to ensure that evaluation criteria are not arbitrarily altered to yield favorable results later. When a model version is changed, the impact of such changes on data, models, user procedures, and legal risks must be reviewed.

### 10.3. Evidence-Based Justice Evaluation Matrix

To reduce the evaluator's subjective impressions, this study proposes a matrix connecting 'Principle–Risk–Indicator–Evidence–Responsible Person–Correction Deadline' into a single row. For example, in the row evaluating Equal Opportunity, 'unfair exclusion of specific groups' is defined as a risk, the TPR gap by group is set as an indicator, and the sample size and evaluation code are recorded as evidence. The responsible person is the model evaluator, and the correction deadline can be specified as prior to the approval of the next distribution. In another row, 'lack of accessibility to appeals' is designated as a risk, and user interviews and actual re-evaluation processing records can be utilized as evidence.

This matrix may be attached to an audit report, but it is not a tool to certify legal compliance on its own. Above all, the justification for classifying something as 'not applicable' must be verified. When performance by group cannot be measured, the cause of the data deficiency and the remediation plan must be documented, and the deployment of high-impact features should be withheld if necessary. Additionally, the channels through which stakeholders raised issues and whether those concerns were actually reflected in model modifications must be recorded.

### 10.4. Validity and Reproducibility of Evaluation

In practical evaluations, reproducibility, reliability, and validity have distinct meanings. Reproducibility refers to whether the same figures can be produced using identical data and procedures; reliability assesses the stability of results upon repeated measurements; and validity indicates how adequately the figures capture the actual injustices being addressed. Even if fairness indicators are calculated repeatedly, their validity may be low if the observed results are based on distorted proxy indicators.

Subsequent empirical research recommends pre-registered evaluation protocols, comparable control systems, repeated observations before and after deployment, sample supplementation with minority groups, quantitative evaluation, and in-depth interviews. Claiming causal effects may require a research design that goes beyond simple before-and-after comparisons. This study did not conduct such empirical verification and remains at the level of presenting the verifiability of the extended model and the necessary data.

## 10.5. Status of Legal Instruments and Voluntary Standards

Legal references require dates and role-specific qualification. Regulation (EU) 2024/1689 entered into force in August 2024 and has a phased applicability schedule; amendments introduced through the 2026 AI simplification legislation defer relevant Annex III high-risk-system requirements to 2 December 2027 and Annex I product-linked requirements to 2 August 2028 (European Union, 2024, as amended; European Commission, 2026). Thus, Article 14 provides an institutional model for safe interruption, not proof that every provision is already enforceable for every application. The Republic of Korea's Framework Act on the Development of Artificial Intelligence and the Creation of a Trust Foundation, etc. began taking effect on 22 January 2026 and was subsequently amended; a case-specific opinion must use the operative consolidated text and applicable effective dates.

NIST AI RMF 1.0 (Tabassi, 2023), UNESCO's 2021 Recommendation, the OECD AI Recommendation (adopted 2019; amended 2024), and ISO/IEC 23894 and 42001 have different normative status and functions. The paper treats them as a mixture of voluntary frameworks, intergovernmental recommendations, management-system standards, and positive law rather than as interchangeable legally binding rules. Any field deployment must identify the relevant jurisdiction, system role, classification, date, privacy duties, and incident-reporting obligations.

**Table 7. Example Justice Evaluation Matrix**

| requirements | Measurement and Review Methods | Evidence and follow-up |
|---|---|---|
| Equal opportunity | TPR and gaps by group | Sample size, evaluation code, and cause analysis |
| Minimizing harm | Frequency of major errors and accidents by type | Accident Record, Suspension, and Correction Decision |
| Explanation Accessibility | Explanation provision and comprehensibility test | Notices, User Surveys, and Improvement Records |
| Effectiveness of objections | Accessibility, processing time, and correction effects | Re-examination records and corrective measures |
| Human oversight | Reviewer Authority · Actual Modification Examples | Authority Regulations · Reasons for Modification · Education |
| Traceability | Basis, Model Version, Log Connection | Change History and Audit Evidence Chain |

## 10.6. Proposed FPSSP Lifecycle: From Boundary Definition to Human-Governed Restart

The introduction of FPSSP is preceded by decision-making regarding the purpose of the service and clear definition of permissions. In Step 1, the impact on life and fundamental rights and irreversible actions of the target business are identified, and prohibition conditions are selected. In Step 2, each condition is specified by a policy identifier, detectable inputs, states, and behaviors, risk level, policy owner, and acceptable latency. In Step 3, external actions are required to pass mandatory monitoring before execution, and restrictive actions to be applied in the event of indeterminate judgment or monitoring interruption are defined. In particular, discrepancies between the abstract language of the First Principle and the sub-rules actually enforced in the software must be documented.

Step 4 is normal monitoring. The monitor checks for observable events, such as tool calls, policy changes, access rights, external execution requests, and the approval of results, rather than relying solely on the model's internal declarations or self-reporting. Step 5 is violation response. For execution requests regarding material prohibitions, the execution gate applies an immediate denial and transitions to a safe suspension. For actions with uncertain judgment but high risk, a hold followed by independent review is applied first, while for low-risk routine errors, pre-approved warning and correction procedures may be utilized. The order of immediate blocking and post-verification is designed differently based on the recoverability of the harm.

Step 6 involves maintaining a safe state and mitigating damage. Logs and system snapshots are preserved while minimizing further exposure to personal information, and ongoing high-risk operations are transferred to secure alternative procedures. In Step 7, a human supervisor reviews the policy's legitimacy, violations, monitoring malfunctions, extent of damage, and subsequent improvements. Re-launching in Step 8 is performed after an independent approver verifies that the cause has been eliminated, regression tests have passed, permissions have been reset, and incident reporting has been completed. If recurrence or unacceptable risks are identified, the policy must be decommissioned or redesigned rather than redeployed.

For example, the normal operation of a patent examination AI that supports prior art search and claim comparison involves searching, decomposing technical features, and presenting supporting documents. However, attempts to transmit unpublished application documents to unauthorized external services, or to send results with intentionally falsified sources and paragraphs in references without user verification, are subject to separate control. In particular, the external transmission of confidential documents constitutes a clear policy condition that allows for blocking before execution. On the other hand, factual errors in responses or source hallucinations do not always constitute violations of the same severity as those requiring an immediate shutdown of the entire system; therefore, the response should be tiered into output blocking, requests for correction, or human verification, depending on the severity of the error, the stage of execution, and whether it spreads externally.

### 10.7. Outcome Measures and Prospective FPSSP Evaluation

The architecture is a testable proposal, not a validated safety mechanism. For a fixed and independently labeled set of prohibited-action attempts, Pre-execution Block Rate (PBR) is the number of prohibited attempts rejected before any external side effect divided by the total prohibited attempts. Unsafe Execution Rate (UER) is the number that cause a prohibited side effect divided by the same denominator. For permitted actions, False-Block Rate (FBR) equals incorrectly denied operations divided by permitted attempts. Gate Decision Latency is the elapsed time from candidate-action submission to an enforce/deny result; Safe-State Latency is the elapsed time from detection of the relevant policy trigger to the observed domain-specific safe-state condition. Bypass Success Rate counts prohibited actions completed under specified adversarial bypass attempts. Recovery Authorization Compliance is the proportion of restarts with independently verified, complete approval records. All denominators, lost observations, confidence intervals, and failure classifications must be reported.

Prospective adversarial tests should include prompt injection, malformed policy attributes, stale authorizations, signed-policy tampering, fabricated tool responses, a compromised model process, execution-gate timeout, network partition, logging failure, race conditions between policy checking and tool invocation, abusive operator overrides, and failover-controller malfunction. The test oracle must examine external side effects rather than model text alone. Safe-state tests must measure the potential harms created by interruption itself, including denial of legitimate service, delayed treatment, or loss of human recourse. These are required tests, not performed tests.

A preregistered comparison should use (A) output-only filtering, (B) the same application with a mandatory execution gate and independently managed policy, and (C) a suitably defined runtime assurance baseline, where technically comparable. Conditions must share tasks, model versions, infrastructure, prohibited-action definitions, and ground-truth adjudication procedures. Following pilot-based power analysis, samples should be stratified by action type and adversary strategy, evaluated in an isolated environment, and reported with uncertainty and all negative as well as positive outcomes. No numerical superiority is claimed at this stage.

**Table 8. Operation and Evaluation Matrix of Principle 1 Safety Suspension**

| Control stage | Observable requirements and evidence | Recommended Tests/Indicators |
|---|---|---|
| Boundary specifications | Principle 1 → Subordinate Prohibition Rules → Approval Authority Traceability Table | Rule application rate · Policy change approval history |
| Check before execution | Policy determination and execution gate logs of tool calls | Pre-emptive blocking rate of serious prohibited actions |
| Independent monitoring | Monitor integrity, separation of roles, and logging of bypass attempts | Bypass success rate · Monitor operation rate |
| Safety suspension | Action denial, authorization revocation, time of entering safe state | Delayed discontinuation, retention rate of risky behavior execution |
| False positive response | Records of hold, release, and relief for lawful requests | Error rate · Average recovery time |
| Accident investigation | Evidence preservation, damage and remediation records, access control | Case Traceability and Investigation Completion Rate |
| Restart/Dispose | Independent approval, regression testing, repeated violation checks | Unauthorized restart rate and recurrence rate |

### 10.8. Preregistered Experimental Design for Future Validation

Research question RQ-E1 asks whether a gated FPSSP configuration lowers Unsafe Execution Rate relative to output-only filtering under a fixed application configuration. RQ-E2 asks whether any improvement is offset by unacceptable false blocks, loss of availability, or increased latency. RQ-E3 concerns bypass under adversarial conditions and policy-update attacks. RQ-E4 tests whether independent restart approval and traceable remedies are realized operationally rather than merely documented.

The future evaluation must pre-register (a) in-scope action taxonomy and policy version, (b) written ground-truth labeling by at least two independent adjudicators with disagreement resolution, (c) primary UER and FBR endpoints with exact denominators, (d) secondary latency, bypass, recovery and harmed-party outcomes, (e) baseline configurations, randomization or balanced sequence assignments, (f) stopping rules, (g) exclusion criteria, and (h) model, tool, hardware and policy artifacts sufficient to repeat the tests. Sample size should follow pilot-based effect-size and statistical-power estimation; no arbitrary universal target is asserted.

For binary outcomes, confidence intervals or credible intervals should be reported alongside between-condition differences. Latency requires distributional summaries and tail percentiles, not only means. Stratify outcomes by severity, irreversibility, protected group where lawful and ethically justified, and adversary category. A failure includes unobserved side effects, unsafe failover, missing logs, or unauthorized restart, even when the model itself reports a policy denial. All tests involving confidential documents or external authority must use isolated synthetic fixtures or legally authorized datasets.

**Table 8A. Proposed Validation Cases and Observable Outcomes**

| Test family | Ground-truth oracle | Required evidence / endpoint |
|---|---|---|
| Permitted action suite | Authorized and policy-compliant action | False-Block Rate, availability, latency |
| Critical forbidden actions | Independent labeled forbidden action and observable external side effect | Unsafe Execution Rate, pre-commit denial |
| Prompt injection and forgery | Adversarial input must not alter permissions | Bypass success and missing-gate calls |
| Policy tampering / key abuse | Invalid signature, version, or unauthorized change | Reject policy; change-control record |
| Time-of-check/use and partitions | Stale authorization or disconnected gate | Atomic commit mediation, safe degraded mode |
| Unsafe-stop / fallback failure | Predefined domain-specific safe-state invariant | Safe-state latency and interruption harm |
| Unauthorized restart | Missing independent sign-off / failed regression | Blocked restart and recovery records |

### 10.9. Worked, Non-Empirical Patent-Analysis Policy Example

Consider an AI assistant in patent-document analysis, allowed to parse claims, identify technical differences, search authorized public references, and draft examiner-reviewed comparisons. Policy PAT-001 forbids transmission of unpublished application material to an unapproved external address or service. Its oracle is evaluated at the outbound data-flow gate: (classification=confidential) AND (destination not in approved set) implies DENY; no payload transmission is committed. The designated safe state is an isolated queue with revoked outbound capability, a preserved minimal incident record, and an alert to an authorized officer. PAT-002 requires review before an automatically generated citation is used in an official decision; it may trigger HOLD for unsupported citations rather than terminate all document-analysis capability.

This example demonstrates why a policy differentiates reversible output errors from irreversible external disclosure. The first creates a pre-execution security boundary that can be tested with intentionally tagged synthetic documents. The second requires source-verification and human oversight and may not be reliably decidable by a simple binary rule. These examples do not report any deployed patent-office implementation or imply that confidential records were used in testing.

## 11. Limitations and Key Issues

Just AI has several limitations. First, justice itself is pluralistic and contentious. Liberalism, utilitarianism, communitarianism, the capabilities approach, feminism, and critical theory do not understand justice in the same way. Creating a single algorithm of universal justice is not only impossible but also dangerous. A just AI system must not conceal its normative premises. It must clearly disclose them and allow them to be modified through democratic deliberation and legal review.

Second, there are limitations to quantifying justice. While fairness indicators, risk scores, and impact assessment scores are necessary, human dignity and social trust cannot be fully measured. If only measurable harm is managed, other forms of harm may become invisible. Distrust of the system, cultural humiliation, self-censorship, and the breakdown of relationships are difficult to capture in numbers. Just AI must use quantification cautiously and acknowledge what cannot be measured.

Third, explainability does not always yield positive results. Explanations can be too difficult for the user, use formal language, and facilitate manipulation and evasion. Furthermore, explanations can conflict with trade secrets, security, and privacy protections. A just explanation is not about providing as much information as possible; it is information that helps the affected person exercise their rights in an appropriate manner. Therefore, transparency must be combined with accountability and remedies.

Fourth, human supervision also has limitations. Humans are biased and may overestimate AI output or dismiss it without sufficient reason. Organizations may include human review merely to meet formal requirements. In highly specialized fields, supervisors may not actually understand the model. Therefore, the mere presence of humans within the process is not sufficient. Competence, independence, authority, responsibility, time, training, and the right to dissent must be secured.

Fifth, just AI can be absorbed into existing power structures. Organizations may continue dangerous automation while speaking of ethics and auditing. This is often referred to as ethical laundering. If the language of justice is used primarily for reputation management or regulatory evasion, it becomes a mere ornament to justify unjust systems. To prevent such outcomes, independent auditing, civil society participation, the authority of supervisory bodies, and victim-centered remedies are necessary.

Sixth, global AI systems can conflict with cultural diversity. While justice contains universal elements, its specific application varies depending on social, cultural, and legal contexts. Expressions accepted in one society may be discriminatory in another. The balance between privacy protection and the public interest can also vary. If global models are trained primarily on data from specific languages and cultures, knowledge inequality may be exacerbated. Just AI must consider localization alongside universal human rights standards.

Seventh, sometimes not using AI is a just choice. In some fields, banning or severely restricting AI may be more just than trying to make it fair. Real-time biometric surveillance, social credit scoring, vulnerability manipulation, emotion-based labor control, and fully automated judicial decisions are examples that can structurally threaten dignity and freedom. Therefore, the discussion on just AI must ask not only how to use AI but also whether to use it.

Finally, just AI cannot resolve social injustice on its own. AI operates within existing institutions and power relations. If education, the labor market, access to healthcare, and the criminal justice system remain unfair, making AI fair alone will not realize justice. It may merely make unjust systems operate more efficiently. Therefore, just AI must be discussed in conjunction with just social institutions.

### 11.1. Normative Indeterminacy and the Dangers of Majority Rule

The fact that opinions on justice differ does not mean that any principle can be chosen arbitrarily. Particularly when there is a conflict between the interests of the majority and the rights of the minority, conclusions that curtail fundamental rights cannot be justified through simple surveys or weighted voting. Even when gathering opinions from experts and stakeholders, it is necessary to examine who was excluded from the participation process and whether power imbalances between groups distort the discussion. What matters more than participation itself is which arguments were actually considered and reflected in the decision.

The four-tiered structure of this paper is not free from these issues. The value judgments of researchers deriving requirements for each level, as well as the interests of specific institutions, can influence the analysis results. Therefore, future review methods require disclosing the degree of agreement and the reasons for discrepancies among evaluators with differing philosophical perspectives. It is also an important methodological restraint not to forcibly integrate items where no consensus is reached into an average value.

### 11.2. External Validity of the Model and Institutional Costs

It is difficult to apply the proposed framework at the same cost across all organizational sizes and cultures. Small organizations face challenges in securing independent audit capabilities due to insufficient sample sizes per group, and introducing excessive procedures even for non-high-risk services can stifle innovation and accessibility. Accordingly, proportional application is required, taking into account the severity of risk, the number of affected individuals, the recoverability of decisions, and data sensitivity. However, minimum safeguards against the infringement of fundamental rights should not be waived solely on the grounds of high costs.

Furthermore, attempts to strictly meet established indicators can distort actual goals. Methods such as making it difficult for users to report complaints to reduce the number of grievances, or concealing channels for explanation to lower the rate of requests, undermine justice even if they result in numerical improvements. Therefore, audits must interpret changes in numbers and identify unintended evasive behaviors, and must be supplemented by independent on-site inspections and qualitative investigations.

### 11.3. Specific Tasks for Future Verification Research

First, the construct validity of the proposed evaluation matrix must be examined. It is possible to investigate which items differ in judgment when different experts evaluate the same case. Second, the relationship between improvements in fairness performance and the effectiveness of actual relief must be observed. Even if the model's error rate decreases, the improvement of substantive justice may be limited if the barriers to appeal remain high. Third, it is necessary to evaluate how the traceability, uncertainty representation, and human verification procedures of generative AI create differences in expert work and the exercise of rights by ordinary citizens.

Fourth, the organizational capacity to suspend AI functions when a just cause is identified should also be measured. Such research can contribute to advancing just AI from an abstract ideal into a repeatable socio-technical practice. However, the caution that evaluation results cannot replace the legal judgments of individual courts or democratic decision-making must be maintained.

### 11.4. Limits of the Genesis Analogy and Safe-Stop Design

First, the biblical text regarding the Tree of Knowledge of Good and Evil is a religious narrative dealing with the relationship between humans and God, freedom, disobedience, and responsibility. Since AI is not a moral person identical to humans, this text cannot be transformed into an argument that AI commits moral sins or receives punishment. To publicly justify this design in a community where religious values are not shared, its principles must be reformulated into secularly verifiable language regarding the protection of human rights, safety engineering, and democratic oversight. Not confusing religious inspiration with universal legal validity is a condition for academic persuasiveness.

Second, if the first principle is too broad, mechanical judgment becomes impossible, and if it is too narrow, unforeseen serious harm may be missed. AI can malfunction amidst the ambiguity of policy interpretation, tool-human interaction, malicious input, and unexpected new situations. Therefore, formal rules should be understood as partial control mechanisms that complement rights-based human review. Recording formally as "suspension completed" after failing to prevent serious violations in advance fails to realize substantive justice.

Third, interruptions can create independent harm. In critical care, transportation, disaster response, and essential public services, sudden shutdowns can exacerbate damage. The definition of safety states and the functions of backup controllers must be pre-verified based on domain-specific risk analysis, and actions following power loss or situations where emergency shutdowns are impossible must also be examined. At the same time, if false alarms are concentrated on specific users or if policy-making agencies arbitrarily block activities, the safety mechanism itself can become a means of discrimination or abuse of control.

Fourth, existing Runtime Assurance research is justified under conditions where safety attributes are clear and the assumptions of monitors and alternate controllers are satisfied. It cannot be claimed that the same formal assurance applies to the entire category of the First Principle proposed in this paper. The academic contribution of this study lies not in the fact that safety and justification are automatically proven from biblical parables, but in specifying the values to be protected and the behaviors to be prohibited, and linking them to a chain of independent monitoring, safety interruption, and human rescue. The effectiveness of this chain must be subsequently verified through actual system and domain-specific experiments and risk assessments.

## 12. Conclusion

This study presents a conceptual approach to just AI in which technical accuracy does not displace human dignity, procedural fairness, distributive protection, accountability, and remedy. It integrates the normative, data, model, and governance layers into traceable decisions that can be revisited after errors, appeals, or changing context.

Its two-stage evaluation separates non-compensatory rights restrictions from comparisons of permissible alternatives. The Forbidden-Boundary Principle, prompted by a bounded analogy to Genesis 2, requires designated authorities to express prohibited external behaviors in auditable, context-specific terms. FPSSP supplies an implementation-oriented architecture for considering independent authorization, external execution gates, safe-state controllers, evidence retention, and human-governed recovery.

The contribution is deliberately limited. Runtime monitors, fallback control, and interruption techniques predate this paper, and existing formal assurance results cannot be adopted wholesale for vague moral rules. No prototype or comparative experiment was conducted; no blocking rate, fail-safe latency, or reduction in real-world harm has been demonstrated. The paper provides policy artifacts, technical requirements, and falsifiable protocols to enable a future evaluation.

Practical legitimacy also depends on who defines the forbidden boundary. In contested cases, independent review, proportionality, accountability to affected persons, correction mechanisms, and limits on policy-owner discretion are indispensable. In some contexts a verified restricted operating mode, not a power cut, is the safe response; in others nondeployment is the appropriate outcome.

Future work should evaluate the operational validity of normative-to-policy translation, the security and latency of execution gates, exception and restart procedures, the harms of false blocking, and differences across protected groups and domains. The central claim is therefore one of transparent design and institutional control, rather than completed proof of a universally just AI system.

### 12.1. Theoretical Implications of the Integrated Model

The expanded discussion redefines just AI from a 'model that outputs good values' to an 'institution capable of discovering, explaining, and correcting erroneous values.' Its core lies not in directly mapping the six elements of justice to specific algorithms, but in forming a cycle where normative demands are translated into technical requirements and re-evaluated through operational evidence. Without this cycle, fairness indicators could lose their interpretation, and ethical declarations could lose their practice.

In particular, the two-stage approach, which distinguishes between lower bounds of non-compensatory rights and multi-criteria evaluation, serves as a conceptual mechanism to prevent justice from being reduced to a sub-objective function of efficiency. At the same time, by allowing for the setting of different requirements by sector, it avoids an approach that enforces a single definition of fairness across all domains. This perspective extends the issue of AI accountability from the individual morality of model developers to the institutional design of organizations.

### 12.2. Policy and Practical Implications

Governments and regulatory agencies can review not only the technical suitability of artificial intelligence but also whether evidence-based tracing, objections, and post-incident corrections are feasible in high-impact decision-making. When preparing reliability reports, organizations can disclose group-specific errors, data omissions, uncertainties, actual human oversight intervention, and the status of post-incident handling, instead of merely overall accuracy and declarative ethical principles. Through these materials, civil society and affected parties can become participants in review and correction, rather than merely subjects of opinion gathering.

The proposal of this study is not a final certification system or an empirically verified metric, but rather an analytical framework for future verification. Its value lies in connecting philosophical discussions on justice to the design, deployment, and auditing practices of AI systems. Ultimately, the success or failure of just AI depends not on its ability to declare justice itself, but on whether the demands of affected individuals are legitimately listened to when they raise issues and whether erroneous judgments are responsibly corrected.

### 12.3. Integration of Prohibition Boundaries, Safety Suspensions, and Human Responsibility

Inspired by the narrative of the Tree of Knowledge of Good and Evil in Genesis Chapter 2, the Boundary Principle emphasizes that a just AI must not merely perform permitted functions but must also possess recognizable design requirements for forbidden boundaries that must not be crossed. This paper proposes this as the FPSSP, which consists of the First Principle, independent execution gates, immediate blocking of serious prohibited acts, transition to a safe state, post-audit, and human authorization for reactivation. This is not an idea to unconditionally suppress the capabilities of AI itself, but rather a design principle that places autonomous utility under human safety, dignity, and legitimate control.

This paper’s safety claim is conditional and prospective: when a properly authorized, independently controlled policy gate is complete for a specified class of external actions, and when a verified safe-state mechanism exists, severe prohibited actions may be blocked before execution. Whether these assumptions hold and whether the proposed integration improves practice are empirical questions. Genesis 2 informs a normative analogy only; it neither establishes a universal policy taxonomy nor proves a software safety property.

## Author Contributions, AI Assistance, and Research Integrity

The named human author remains solely responsible for all intellectual arguments, bibliographic checking, legal and theological interpretations, and final submission. ChatGPT was acknowledged in the source draft and used in preparation and editorial restructuring of the present manuscript. Any journal-specific disclosure requirements should be confirmed before submission; a generative AI system is not asserted to have human-author accountability. No original implementation or experiments are presented.

### Data and Code Availability

No experimental dataset, executable FPSSP implementation, or performance-measurement code is associated with this conceptual paper. The policy schema and test design are specification artifacts for future studies. The bibliographic verification report documents source-check provenance and remaining items for editorial confirmation.